\documentclass[a4paper,11pt]{article}
\usepackage[margin=2.5cm]{geometry}
\usepackage[utf8]{inputenc}
\usepackage{graphicx}
\usepackage{epsfig}
\usepackage{amsmath}
\usepackage{amssymb}
\usepackage{cite}
\usepackage{float}
\usepackage{subfigure}
\usepackage[font={footnotesize,it}]{caption}
\usepackage{authblk}
\numberwithin{equation}{section}
\usepackage{hyperref}
\hypersetup{
    colorlinks=true,
    linkcolor=blue,
    filecolor=red,      
    urlcolor=blue,
    citecolor=red
} 
\usepackage{epstopdf}
\DeclareGraphicsExtensions{.eps}
\usepackage{braket}
\usepackage{tikz}

\begin{document}

\begin{titlepage}

\title{Covariant Holographic Entanglement Negativity for Adjacent Subsystems}
\author[1,2,3]{Parul Jain\thanks{\noindent E-mail:~ parul.jain@ca.infn.it}}
\affil[1]{Dipartimento di Fisica, Universit\`a di Cagliari\\Cittadella Universitaria, 09042 Monserrato, Italy\smallskip}
\affil[2]{INFN, Sezione di Cagliari, Italy\bigskip}
\author[3]{Vinay Malvimat\thanks{\noindent E-mail:~ vinaymm@iitk.ac.in}} 
\author[3]{Sayid Mondal\thanks{\noindent E-mail:~  sayidphy@iitk.ac.in}}
\author[3]{Gautam Sengupta\thanks{\noindent E-mail:~  sengupta@iitk.ac.in}}
\affil[3]{Department of Physics\\

Indian Institute of Technology\\
 
Kanpur, 208016\\ 

India}

\maketitle

\begin{abstract}
We propose a covariant holographic entanglement negativity construction for time dependent mixed states of adjacent intervals in $(1+1)$-dimensional conformal field theories dual to non-static bulk $\mathrm{AdS_3}$ configurations. Our construction applied to $(1+1)$-dimensional conformal field theories with a conserved charge dual to non-extremal and extremal rotating BTZ black holes exactly reproduces the corresponding replica technique results, in the large central charge limit. We also utilize our construction to obtain the time dependent holographic entanglement negativity for the mixed state in question for the $(1+1)$- dimensional CFT dual to a bulk Vaidya-$\mathrm{AdS_3}$ configuration.

\end{abstract}
\end{titlepage}
\tableofcontents

\section{Introduction}

Quantum entanglement is one of the most intriguing aspects of quantum information theory which still eludes a complete understanding. Recent advances in this area
have led to significant implications across diverse fields from condensed matter physics to quantum gravity. In this context {\it entanglement entropy} has emerged as a crucial measure to characterize the entanglement of bipartite quantum systems in pure states.
This measure is defined as the von Neumann entropy for the reduced density matrix of the corresponding subsystem under consideration. Although it is straightforward to compute this quantity for systems with finite degrees of freedom, it is a complex and subtle issue
for extended quantum many body systems. Despite this complication, a {\it replica technique} allows a systematic approach for the computation of this quantity in $(1+1)$-dimensional conformal field theories $(\mathrm{CFT}_{1+1})$ \cite{Calabrese:2004eu,Calabrese:2009qy}.

It is well known in quantum information theory that the entanglement entropy is a valid measure for characterizing the entanglement of bipartite systems in pure states only. The issue of mixed state entanglement on the other hand is still an active area of research in the subject. For such mixed quantum states the entanglement entropy receives contributions from correlations which are not relevant for characterizing the entanglement of the bipartite system in question. In a classic communication \cite{PhysRevA.65.032314}, Vidal and Werner addressed this significant issue and proposed  a computable measure termed {\it entanglement negativity} which led to an upper bound on the distillable entanglement. Interestingly in \cite{Calabrese:2012ew,Calabrese:2012nk,Calabrese:2014yza}, the authors computed the entanglement negativity for various mixed states in $\mathrm{CFT}_{1+1}$ employing a variant of the replica technique mentioned earlier. 

In the context of the $\mathrm{AdS/CFT}$ correspondence Ryu and Takayanagi (RT) \cite{Ryu:2006bv,Ryu:2006ef} proposed an elegant holographic conjecture for the
entanglement entropy of $d$-dimensional boundary CFTs ($\mathrm{CFT_d}$) dual to bulk $\mathrm{AdS_{d+1}}$ configurations. The holographic entanglement entropy $S_{\gamma}$ in this instance could be described in terms of the
area ${\cal A}_{\gamma}$ of a co-dimension two bulk static minimal surface anchored on the subsystem $\gamma$ under consideration as follows 
\begin{equation}\label{RT}
S_\gamma=\frac{{\cal A}_{\gamma}}{4~G^{(d+1)}_N},
\end{equation}
where $G^{(d+1)}_N$ is the $(d+1)$-dimensional Newton constant. Since its proposal the Ryu and Takayanagi conjecture has led to intensive research  providing significant insights into the connections between space time geometry and entanglement \cite {Takayanagi:2012kg,Nishioka:2009un,1751-8121-42-50-500301,Cadoni:2009tk,e12112244,Hubeny:2012ry,Fischler:2012ca,Fischler:2012uv,PhysRevD.94.066004}.

The Ryu-Takayanagi \cite{Ryu:2006bv,Ryu:2006ef} conjecture described above refers to the time independent entanglement in holographic CFTs which are dual to bulk static AdS configurations. For time dependent entanglement scenarios it is necessary to determine
the precise definition of minimal surfaces for spacetime geometries with a Lorentzian signature. For such Lorentzian signature it is possible to contract a spacelike surface in the time direction to make its area arbitrarily small. This issue was resolved in a significant communication by Hubeny, Rangamani and Takayanagi (HRT) in which they proposed a covariant holographic conjecture for the entanglement entropy of CFTs dual to non-static bulk AdS gravitational configurations \cite{Hubeny:2007xt}. Their conjecture involves the light sheet construction arising from the covariant Bousso entropy bound \cite{Bousso:1999xy,Bousso:1999cb,Bousso:2002ju} to describe the time dependent holographic entanglement entropy $S_{A_t}$ of a subsystem $A_t$ in terms of the area of co-dimension two bulk extremal surface $ {\cal Y}_{A_t}^{ext}$ as follows
\begin{equation}\label{ententropyrt}
 S_{A_t}=\frac{\mathrm{Area}~({\cal Y}_{A_t}^{ext})}{4G_N^{(d+1)}}.
\end{equation}
This elegant covariant HRT conjecture has been extensively employed to investigate various time dependent entanglement issues for holographic CFTs \cite {PhysRevD.84.026010,Hartman:2013qma,1367-2630-13-4-045017,Caputa:2013eka,Mandal:2016cdw,David2016}.

As stated earlier the entanglement entropy fails to be a valid measure
for characterizing mixed state entanglement in quantum information. Thus the holographic entanglement entropy conjectures of RT and HRT are applicable only for characterizing pure state entanglement in holographic CFTs. In the recent past two of the present authors (VM and GS) in the collaborations \cite {Chaturvedi:2016rcn,Chaturvedi:2016rft} proposed an elegant conjecture for the entanglement negativity of holographic CFTs (see also the discussions in \cite{Rangamani:2014ywa} and for recent applications see 
\cite{Erdmenger:2017pfh}). A covariant entanglement negativity construction for time dependent mixed state entanglement in holographic CFTs was subsequently proposed in \cite {Chaturvedi:2016opa}.

Note that the entanglement negativity conjecture in 
\cite {Chaturvedi:2016rcn,Chaturvedi:2016rft} refers to the mixed state entanglement between a single subsystem with the rest of the system in a holographic CFT. Motivated by the above construction very recently the present authors have proposed a holographic entanglement negativity construction relevant for mixed states of adjacent subsystems in a holographic CFT \cite {Jain:2017aqk,Jain:2017xsu}. Remarkably in the 
$\mathrm{AdS_3/CFT_2}$ context \cite {Jain:2017aqk} this independent construction exactly reproduces the entanglement negativity of a holographic $\mathrm{\mathrm{CFT_{1+1}}}$ obtained through the replica technique,
\cite {Calabrese:2012ew,Calabrese:2012nk} in the large central charge limit \cite {Kulaxizi:2014nma}.

In this article we propose a covariant holographic construction to describe the time dependent entanglement negativity for mixed states of adjacent intervals in holographic $\mathrm{\mathrm{CFT_{1+1}}}$s dual to non-static bulk $\mathrm{AdS_3}$ configurations. This analysis in the $\mathrm{AdS_3/CFT_2}$ framework is expected to provide crucial insight into the corresponding higher dimensional construction for the holographic entanglement negativity in a generic $AdS_{d+1}/CFT_d$ scenario. Our holographic construction for the entanglement negativity in the $\mathrm{AdS_3/CFT_2}$ scenario described above involves a specific algebraic sum of the lengths of extremal curves anchored on appropriate combinations of intervals in the boundary $\mathrm{CFT_{1+1}}$. This may be expressed as an algebraic sum of the corresponding holographic entanglement entropies for these combinations of the intervals from the HRT conjecture \footnote{Interestingly, the algebraic sum of the holographic entanglement entropies reduce to the holographic mutual information between the adjacent intervals up to a numerical factor. This has been observed for global and local quench scenario in \cite{Coser:2014gsa,Wen:2015qwa} respectively}. We employ our holographic construction for the 
$\mathrm{AdS_3/CFT_2}$ scenario to compute the entanglement negativity of such mixed states in $\mathrm{\mathrm{CFT_{1+1}}}$s with a conserved charge which are dual to bulk non-extremal and extremal rotating BTZ black holes. Quite remarkably, our results exactly match with the entanglement negativity of such holographic $\mathrm{\mathrm{CFT_{1+1}}}$s computed through the replica technique, in the large central charge limit. Interestingly, we observe that the covariant entanglement negativity
for the holographic $\mathrm{\mathrm{CFT_{1+1}}}$s dual to non-extremal rotating BTZ black holes decomposes into left and right moving sectors with distinct temperatures as described in \cite {Zoltan}. Curiously, for the extremal rotating BTZ black holes, the holographic entanglement negativity characterized by a bulk Frolov-Thorne temperature \cite {PhysRevD.39.2125}, matches exactly with that obtained from the chiral half of the dual 
$\mathrm{\mathrm{CFT_{1+1}}}$. Note that the above bulk configurations are stationary but not time dependent. 

As a further consistency check for our covariant holographic negativity construction, we further compute the time dependent negativity for the mixed state described above in a $\mathrm{\mathrm{CFT_{1+1}}}$ dual to a bulk Vaidya-$\mathrm{AdS_3}$ configuration. Very interestingly we observe numerically that the holographic negativity monotonically decreases with time clearly indicating the thermalization of the mixed state in question which corresponds to the black hole formation in the dual bulk $\mathrm{AdS_3}$ geometry.

This article is organized as follows. In section \ref{sec_rev_neg} we briefly review the entanglement negativity in a $\mathrm{CFT}_{1+1}$. In section \ref {sec_rev_HRT} we very briefly recollect the essential elements of the HRT conjecture.  Subsequently in section \ref{sec_cohen_2adj} we describe our covariant holographic entanglement negativity construction. In section \ref{sec_rot_btz} we employ our construction to compute the covariant entanglement negativity for mixed states of adjacent intervals in holographic $\mathrm{\mathrm{CFT_{1+1}}}$s dual to both non-extremal and extremal rotating BTZ black holes. In the subsequent section \ref{sec_CFT_rot_btz} we describe the computation for the entanglement negativity in the dual $\mathrm{\mathrm{CFT_{1+1}}}$s for both the above cases, through the replica technique and compare these with the corresponding holographic results. In section \ref{sec_tdhen} we compute the time dependent holographic entanglement negativity for the mixed state in question, in a $\mathrm{\mathrm{CFT_{1+1}}}$ dual to a bulk Vaidya-$\mathrm{AdS_3}$ configuration and analyse its time evolution. In the final section \ref{sec_sumandcon} we summarize our results and conclusions and describe interesting future open issues.
%%%%%%%%%%%%%%%%%%%%%%%%%%%%%%%%%%%%%%%%Review of HRT%%%%%%%%%%%%%%%%%%%%%%%%%%%%%%%%%%%%%%%%%%%%
%%%%%%%%%%%%%%%%%%%%%%%%%%%%%%%%%%%%%%%%%%%%%%%%%%%%%%%%%%%%%%%%%%%%%%%%%%%%%%%%%%%%%%%%%%%%%%%%
%%%%%%%%%%%%%%%%%%%%%%%%%%%%%%%%%%%%%%%%%%%%%%%%%%%%%%%%%%%%%%%%%%%%%%%%%%%%%%%%%%%%%%%%%%%%%%%%
\section{Review of entanglement negativity}\label{sec_rev_neg}

\begin{figure}[h]
\includegraphics[scale=1.25]{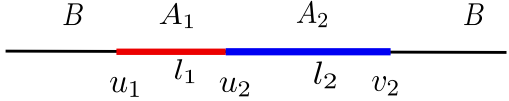}
\caption{Schematic of two adjacent intervals $A_1$ and $A_2$ in a $(1 + 1)$-dimensional CFT.}\label{adjint}
\end{figure}

In this section we briefly review the entanglement negativity for mixed states of two adjacent intervals as depicted in Fig. \ref{adjint} in a $\mathrm{CFT_{1+1}}$ \cite{Calabrese:2012ew,Calabrese:2012nk,Calabrese:2014yza}. In quantum information theory the entanglement between the intervals $A_1$ and $A_2$ of a bipartite system $A=A_1\cup A_2$ in a mixed state is characterized by the entanglement negativity, which is defined as \cite{PhysRevA.65.032314}
\begin{equation}\label{ent_neg}
\mathcal{E} = \ln \mathrm{Tr}|\rho_A^{T_2}|.
\end{equation}
Here the trace norm $\mathrm{Tr}|\rho_A^{T_2}|$ is given by the sum of absolute eigenvalues of the partially transposed reduced density matrix $\rho_A^{T_2}$ which is defined as follows
\begin{equation}\label{26}
\langle e^{(1)}_ie^{(2)}_j|\rho_A^{T_2}|e^{(1)}_ke^{(2)}_l\rangle = 
\langle e^{(1)}_ie^{(2)}_l|\rho_A|e^{(1)}_ke^{(2)}_j\rangle,
\end{equation}
where $|e^{(1)}_i\rangle$ and $|e^{(2)}_j\rangle$ are the bases for the the Hilbert spaces $\mathcal{H}_{A_1}$ and  $\mathcal{H}_{A_2}$ of the subsystems $A_1$ and $A_2$ respectively. In the context of a $\mathrm{CFT_{1+1}}$ the entanglement negativity is obtained through a suitable replica technique by constructing  the quantity $\mathrm{Tr} \big(\,\rho_A^{T_{2}}\big)^{n}$ for even $n=n_e$ and considering the analytic continuation to $n_e\to 1$ as follows \cite{Calabrese:2012ew,Calabrese:2012nk,Calabrese:2014yza}
\begin{equation}\label{en_replica}
\mathcal{E} = \lim_{n_e \rightarrow 1}  \ln \Big[ \mathrm{Tr} \big(\,\rho_A^{T_{2}}\big)^{n_e} \Big]\,.
\end{equation}

\section{Review of the HRT conjecture}\label{sec_rev_HRT}
In this section we very briefly recollect the essential elements of the HRT conjecture \cite{Hubeny:2007xt} mentioned in the introduction. Their covariant holographic entanglement entropy conjecture crucially incorporates the light-sheet construction for the covariant Bousso entropy bound \cite{Bousso:1999xy,Bousso:1999cb,Bousso:2002ju}. A light-sheet $L_{{\cal S}}$ for a co-dimension two space like surface ${\cal S}$ in a space time manifold $\cal M$ is described by the congruence of null geodesics whose expansion is non-positive definite ($\theta_{\pm} \leq 0$, where $\theta_{\pm}$ are the future and the past null expansions respectively). The area of this surface ${\cal S}$ serves as a covariant bound on the entropy flux $S_{L_{\cal S}}$ (in Planck units) crossing through the light sheets as follows
\begin{equation}\label{entropybound}
 S_{L_{\cal S}}\leq \frac{\mathrm{Area}~({\cal S})}{4G_N}.
\end{equation}

According to the HRT conjecture, the holographic entanglement entropy of a subsystem in a 
$\mathrm{CFT_d}$ dual to a bulk $AdS_{d+1}$ configuration, saturates the covariant Bousso bound described above. To this end the $d$-dimensional boundary of the bulk $AdS_{d+1}$ configuration is partitioned into two space like regions $A_t$ and its complement $ A_t^c$ at a constant time $t$. The common $(d-2)$-dimensional boundary shared by these two space like regions is denoted as $\partial  A_t$. The future and the past light sheets $\partial L_{\pm}$ for this space like surface $\partial A_t$ are considered. These light sheets on the $AdS$ boundary may then be extended into the bulk as $L_-$ and $L_+$  where they become the light sheets for a $(d-1)$-dimensional space like surface ${\cal Y}_{A_t}=L_+\cap L_-$ anchored on the subsystem $A_t$. The holographic proposal by HRT states that out of all such  surfaces ${\cal Y}_{A_t}$, the entanglement entropy is determined by the surface which has minimal area $({\cal Y}_{A_t}^{min})$. Furthermore, the authors also demonstrated that the surface  $({\cal Y}_{A_t}^{min})$ was actually the extremal surface (${\cal Y}_{A_t}^{ext}$) anchored on the subsystem $ A_t$, with vanishing null expansions $(\theta_{\pm}=0)$ leading to the saturation of the Bousso bound. The holographic entanglement entropy for the subsystem $ A_t$ is therefore given as
\begin{equation}\label{ententropy}
 S_{A_t}=\frac{\mathrm{Area}~({\cal Y}_{A_t}^{min})}{4G_N^{(d+1)}}=\frac{\mathrm{Area}~({\cal Y}_{A_t}^{ext})}{4G_N^{(d+1)}}.
\end{equation}

Having presented the essential elements of the HRT conjecture, in the next section we describe our covariant holographic construction for the entanglement negativity of mixed states described by adjacent subsystems in a boundary $\mathrm{\mathrm{CFT_{1+1}}}$. Note that for the $\mathrm{AdS_3/CFT_2}$ scenario as mentioned in the Introduction the covariant holographic entanglement entropy involves the length of the extremal curve anchored 
on the corresponding interval which are essentially geodesics.

%%%%%%%%%%%%%%%%%%%%%%%%%%%%%%%%%%%%%%%%%%%%%%%%%%%%%%%%%%%%%%%%%%%%%%%%%%%%%%%%%%%%%%%%%%%%%%%%        Covariant Holographic Entanglement Negativity Conjecture for Two Adjacent Intervals %%%%%%%%%%%%%%%%%%%%%%%%%%%%%%%%%%%%%%%%%%%%%%%%%%%%%%%%%%%%%%%%%%%%%%%%%%%%%%%%%%%%%%%%%%%%%%%%%%%%%%%%%%%%%%
\section{Covariant holographic entanglement negativity for two adjacent subsystems }\label{sec_cohen_2adj}

Here we describe our covariant holographic construction for the time dependent entanglement negativity of the mixed state of adjacent intervals $A_1$ and $A_2$ of lengths $l_1$ and $l_2$ respectively in a $\mathrm{\mathrm{CFT_{1+1}}}$ in the context of the $AdS_3/CFT2$ scenario. To this end we begin with a brief description of our earlier holographic entanglement negativity construction for the corresponding time independent scenario in \cite{Jain:2017aqk} in a $\mathrm{\mathrm{CFT_{1+1}}}$. For this mixed state configuration the entanglement negativity involves the three point twist correlator whose form is fixed from conformal invariance as follows
\begin{equation}\label{threepoint}
\begin{split}
&\langle\mathcal{T}_{n_e}(z_1)\overline{\mathcal{T}}_{n_e}^{2}(z_2)\mathcal{T}_{n_e}(z_3)\rangle_{\mathbb{C}} \\&~~~~~=  
\frac{c_n^2~C_{\mathcal{T}_{n_e}\overline{\mathcal{T}}_{n_e}^2\mathcal{T}_{n_e}}}{|z_{12}|^{\Delta_{\mathcal{T}^2_{n_e}}}
|z_{23}|^{\Delta_{\mathcal{T}^2_{n_e}}}|z_{13}|^{2\Delta_{\mathcal{T}_{n_e}}-\Delta_{{\mathcal{T}}^2_{n_e}}}},
\end{split}
\end{equation}
where $c_n$ are normalization constants and $C_{\mathcal{T}_{n_e}\overline{\mathcal{T}}_{n_e}^2\mathcal{T}_{n_e}}$ is the structure constant for the relevant twist field OPEs. The scaling dimensions $\Delta_{\mathcal{T}_{n_e}}$ and 
$\Delta_{\mathcal{T}_{n_e}^2}$ of the twist fields $\mathcal{T}_{n_e}$ and $\mathcal{T}_{n_e}^2$ are respectively given as
\begin{equation}\label{34}
\begin{aligned}
\Delta_{\mathcal{T}_{n_e}} =  \frac{c}{12}\Big(n_e - \frac{1}{n_e}\Big),\\
\Delta_{\mathcal{T}_{n_e}^2} =2\Delta_{\mathcal{T}_{\frac{n_e}{2}}}= \frac{c}{6}\Big(\frac{n_e}{2}-
\end{aligned} \frac{2}{n_e}\Big).
\end{equation}

Note that in the $\mathrm{AdS_3/CFT_2}$ scenario, it was shown in \cite{Jain:2017aqk} that the universal part of the three point twist correlator related to the entanglement negativity is dominant in the large central charge limit and factorizes into a specific combination of the two point correlation functions which is given as
\begin{equation}
\begin{split}
&\langle\mathcal{T}_{n_e}(z_1)\overline{\mathcal{T}}_{n_e}^{2}(z_2)\mathcal{T}_{n_e}(z_3)\rangle_{\mathbb{C}}=\big<{\cal T}_{n_e}(z_1)\overline{{\cal T}}_{n_e}(z_3)\big>_{\mathbb{C}}\\&~~~~~~\frac{\big<{\cal T}_{\frac{n_e}{2}}(z_1)\overline{{\cal T}}_{\frac{n_e}{2}}(z_2)\big>_{\mathbb{C}}\big<{\cal T}_{\frac{n_e}{2}}(z_2)\overline{{\cal T}}_{\frac{n_e}{2}}(z_3)\big>_{\mathbb{C}}}{\big<{\cal T}_{\frac{n_e}{2}}(z_1)\overline{{\cal T}}_{\frac{n_e}{2}}(z_3)\big>_{\mathbb{C}}}+O[\frac{1}{c}].\label{factor}
\end{split}
\end{equation}
Here we have also used the following factorization of the two point twist correlator in a $\mathrm{\mathrm{CFT_{1+1}}}$ as described in \cite{Calabrese:2012ew,Calabrese:2012nk} 
\begin{equation}
 \langle{\cal T}^2_{n_e}(z_i)\overline{{\cal T}}^2_{n_e}(z_j)\rangle_{\mathbb{C}}=\big<{\cal T}_{\frac{n_e}{2}}(z_i)\overline{{\cal T}}_{\frac{n_e}{2}}(z_j)\big>^2_{\mathbb{C}} .
\end{equation}

\begin{figure}[h]
\includegraphics[scale=1.25]{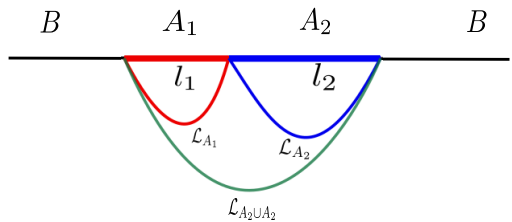}
\caption{Schematic of bulk geodesics anchored on the subsystems $A_1,~A_2$ and $A_1 \cup A_2$ in a $(1 + 1)$-dimensional boundary CFT.}\label{geo_struc}
\end{figure}

Now from the $AdS_3/CFT_2$ dictionary the two point twist correlator is related to the length of the dual bulk space like geodesic 
$\mathcal{L}_{ij}$ anchored on the corresponding intervals as follows \cite{Ryu:2006bv,Ryu:2006ef}
\begin{eqnarray}
&\big<{\cal T}_{n_e}(z_k)\overline{{\cal T}}_{n_e}(z_l)\big>_{\mathbb{C}} \sim e^{-\frac{\Delta_{n_e}{ \cal L}_{kl}}{R}},\label{4}\\
&\big<{\cal T}_{\frac{n_e}{2}}(z_i)\overline{{\cal T}}_{\frac{n_e}{2}}(z_j)\big>_{\mathbb{C}} \sim e^{-\frac{\Delta_{\frac{n_e}{2}}{ \cal L}_{ij}}{R}},~~~~~\label{5}
\end{eqnarray}
where $R$ is the $AdS_3$ radius. The three point twist correlator in eq. (\ref{factor}), upon utilizing eqs. \eqref{4} and \eqref{5}, may be expressed as follows
\begin{equation}\label{3ptfact}
\begin{split}
 &\langle\mathcal{T}_{n_e}(z_1)\overline{\mathcal{T}}_{n_e}^{2}(z_2)\mathcal{T}_{n_e}(z_3)\rangle_{\mathbb{C}}\\& ~~~
=\exp{\bigg[\frac{-\Delta_{\mathcal{T}_{n_e}}\mathcal{L}_{13}-\Delta_{\mathcal{T}_{\frac{n_e}{2}}}(\mathcal{L}_{12}+\mathcal{L}_{23}-\mathcal{L}_{13})}{R}\bigg]}.
\end{split}
\end{equation}
The holographic entanglement negativity of the mixed state of adjacent intervals may then be obtained by utilizing the eqs. \eqref{3ptfact} and \eqref{en_replica}, which is given as a specific algebraic sum of the lengths of the geodesics $\mathcal{L}_{ij}$ anchored on the intervals as depicted in Fig. \ref{geo_struc} as follows
\begin{equation}\label{HEN CONJ AREA}
\mathcal{E} = \frac{3}{16G^{(3)}_N}\big(\mathcal{L}_{A_1}+\mathcal{L}_{A_2}-\mathcal{L}_{A_{1}\cup A_2}\big).
\end{equation}

For the time dependent scenario the covariant HRT conjecture \cite{Hubeny:2007xt} proposes that the entanglement entropy is described by the corresponding area of the co-dimension two bulk extremal surface anchored on the respective subsystem instead of the static minimal surface. As described in the previous section for the $\mathrm{AdS_3/CFT_2}$ scenario the HRT conjecture for the entanglement entropy involves  the length of the bulk extremal curve anchored on the corresponding interval. This motivates us to advance a holographic construction for the time dependent entanglement negativity for the mixed state of adjacent intervals in the $\mathrm{AdS_3/CFT_2}$ scenario as follows 
\begin{equation}\label{COHEN CONJ}
  {\cal E} = \frac{3}{16G_{N}^{(3)}} \Big[{\cal L}_{A_1^t}^{ext}+{\cal L}_{A_2^t}^{ext}-{\cal L}^{ext}_{A_1^t\cup A_2^t}\Big].
\end{equation}
Here ${\cal L}_{\gamma}^{ext}$ ($\gamma \in A_1^t, A_2^t, A_{1}^t\cup A_2^t $) are lengths of extremal curves in the bulk anchored on the respective subsystems on the boundary. It may be shown using the HRT conjecture that the above expression reduces to the holographic mutual information between the adjacent subsystems as follows
\begin{equation}
\begin{aligned}
{\cal E}=\frac{3}{4}\Big(S_{A_1^t}+S_{A_2^t}-S_{A_{1}^t\cup A_2^t}\Big)=\frac{3}{4}{\cal I}(A_1^t,A_2^t),
\end{aligned}
\end{equation}
where $S_{\gamma}$ is the entanglement entropy of a subsystem $\gamma$ Eq. (\ref{ententropy}). It is interesting  to note that the covariant holographic entanglement negativity for adjacent intervals is exactly proportional to the holographic mutual information for the mixed state configuration in question. Note that the preceding proportionality between the two quantities refer only to their universal parts which are dominant in the large central charge limit. The two quantities including their non universal contributions are however not proportional to each other \footnote {Note that the exact proportionality between the holographic entanglement negativity and the holographic mutual information is specific to the adjacent interval configuration. For disjoint intervals however this is not valid as the entanglement negativity goes to zero beyond a certain separation of the intervals in a $\mathrm{\mathrm{CFT_{1+1}}}$ \cite{Calabrese:2012nk}.}. This conforms with the fact that they constitute distinct measures in quantum information theory. Significantly for the same configuration it was shown by Coser et. al. in \cite {Coser:2014gsa} and by Wen et. al. in \cite {Wen:2015qwa}, that the universal parts of these quantities are also proportional in the context of global and local quenches in a $\mathrm{\mathrm{CFT_{1+1}}}$.

%%%%%%%%%%%%%%%%%%%%%%%%%%%%%%%%%%%%%%%%%%%%%%%%%%%%%%%%%%%%%%%%%%%%%%%%%%%%%%%%%%%%%%%%%%%%%%%%
%%%%%%%Covariant Holographic Entanglement Negativity of rotating BTZ Black Holes%%%%%%%%%%%%%%%%%%%%%%%%%%%%%%%%%%%%%%%%%%%%%%%%%%%%%%%%%%%%%%%%%%%%%%%%%%%%%%%%%%%%%%%%%
%%%%%%%%%%%%%%%%%%%%%%%%%%%%%%%%%%%%%%%%%%%%%%%%%%%%%%%%%%%%%%%%%%%%%%%%%%%%%%%%%%%%%%%%%%%%%%%%
\section{ Rotating BTZ black hole}\label{sec_rot_btz}
In this section we utilize our covariant holographic negativity construction described above to compute the entanglement negativity of mixed states of adjacent intervals in holographic $\mathrm{\mathrm{\mathrm{CFT_{1+1}}}}$s with a conserved charge, dual to bulk non-extremal and extremal rotating BTZ black holes.  Note that in this scenario the bulk dual configurations are stationary ( non-static) and the corresponding mixed states of adjacent intervals in the$\mathrm{\mathrm{CFT_{1+1}}}$ are time independent. However these examples constitute significant consistency checks for our construction in the context of the $AdS_3/CFT_2$ scenario.

%%%%%%%%%%%%%%%%%%%%%%%%%%%%%%%%%%%%%%%%%%%%%%%%%%%%%%%%%%%%%%%%%%%%%%%%%%%%%%%%%%%%%%%%%%%%%%%%
%%%%%%%%%%%%%%%%%%%%%%%%%%%%%%%%%%%%%%%%%%%%%%%%%%%%%%%%%%%%%%%%%%%%%%%%%%%%%%%%%%%%%%%%%%%%%%%%
\subsection{Non-extremal rotating BTZ black hole}
The non-extremal rotating BTZ (black string) metric with the AdS length scale set to unity $(R=1)$  is given as
\begin{equation}\label{nonextbtz}
\begin{aligned}
 ds^{2}&=-\frac{(r^2-r_{+}^2)(r^2-r_{-}^2)}{r^2}dt^2 \\
 &+ \frac{r^2 dr^2}{(r^2-r_{+}^2)(r^2-r_{-}^2)}+ r^2(d\phi-\frac{r_{+}r_{-}}{r^2}dt)^2,
 \end{aligned}
\end{equation}
where the coordinate $\phi$ is non compact and $r_{\pm}$ are the radii of the inner and outer horizons respectively. The mass $M$, angular momentum $J$, Hawking temperature $T_H$, and the angular velocity $\Omega$ are related to the horizon radii $r_{\pm}$ as follows
\begin{eqnarray}\label{MJTOmega}
 M = r_{+}^2+r_{-}^2,~~~~~~~~ J=2 r_{+}r_{-},\\
  T_H=\frac{1}{\beta}=\frac{r_{+}^2-r_{-}^2}{2\pi r_+},  ~~~~~\Omega=\frac{r_{-}}{r_{+}}.
\end{eqnarray}
The corresponding dual finite temperature $\mathrm{\mathrm{\mathrm{CFT_{1+1}}}}$ with the bulk angular momentum $J$ as the conserved charge, is defined on a twisted infinite cylinder \footnote {The fact that dual $\mathrm{\mathrm{\mathrm{CFT_{1+1}}}}$ lives on a twisted cylinder can be observed from the conformal transformation given in eq.(\ref{non-extmaps}).}. The corresponding temperatures relevant to the left and the right moving modes in the dual CFT defined on the cylinder may now be expressed as follows
\begin{equation}\label{cfttemp1}
T_{\pm}=\frac{1}{\beta_{\pm}}=\frac{r_+\mp r_-}{2\pi}.
\end{equation}
The covariant holographic entanglement entropy $S_{\gamma}$ for a single spacelike interval $\gamma$ at a constant time $t=t_0$ in the finite temperature holographic $\mathrm{\mathrm{\mathrm{CFT_{1+1}}}}$ dual to the non-extremal rotating BTZ black hole is obtained from the HRT conjecture \cite{Hubeny:2007xt}. To this end it is necessary to perform the following coordinate transformations
\begin{equation}\label{nonextmap}
\begin{aligned}
 w_{\pm}=&\sqrt{\frac{r^2-r_{+}^2}{r^2-r_{-}^2}} e^{2\pi T_{\pm}u_{\pm}}\equiv X\pm T,\\
  Z=&\sqrt{\frac{r_{+}^2-r_{-}^2}{r^2-r_{-}^2}} e^{\pi T_+ u_+ + \pi T_- u_- },
\end{aligned}
\end{equation}
where $u_{\pm}=\phi \pm t$. These transformations map the BTZ metric Eq. (\ref{nonextbtz}) to the $\mathrm{AdS_3}$ metric in the Poincar\'{e} coordinates as
\begin{equation}\label{poincare}
 ds^2=\frac{dw_{+}dw_{-}+dZ^2}{Z^2}\equiv\frac{-dT^2+dX^2+dZ^2}{Z^2}.
\end{equation}
The corresponding length ${\cal L}_{\gamma}$ of the geodesic anchored on the interval 
$\gamma$ of length $\Delta\phi =|\phi_1- \phi_2|$ in the holographic $\mathrm{\mathrm{\mathrm{CFT_{1+1}}}}$ is then given as \cite{Hubeny:2007xt}
\begin{equation}\label{nonextlenght}
\begin{aligned}
{\cal L}_{\gamma}&=\log\bigg[\frac{(\Delta \phi)^2}{\varepsilon_{i}\varepsilon_{j}}\bigg],\\
\varepsilon_{i} &= \sqrt{\frac{r_{+}^2-r_{-}^2}{r_{\infty}^2}}e^{(r_{+}\phi_{i}~-~t_0 r_{-})}.
\end{aligned}
\end{equation}
Here $r_{\infty}$ and $\varepsilon_{i}$ ($i=1,2$) are the infra red cut-off for the bulk in the BTZ and the Poincar\'{e} coordinates respectively. This leads to the following expression for the 
geodesic length ${\cal L}_{\gamma}$ in terms of the original BTZ coordinates as
follows \cite {Hubeny:2007xt}
\begin{equation}\label{s36}
  {\cal L}_{\gamma} =\log\bigg[\frac{\beta_{+}\beta_{-}}{\pi^2a^2}\sinh{\big(\frac{\pi \Delta \phi}{\beta_{+}}\big)}\sinh{\big(\frac{\pi\Delta \phi}{\beta_{-}}\big)}\bigg],
\end{equation}
where $a$ is the UV cut-off for the boundary $\mathrm{\mathrm{\mathrm{CFT_{1+1}}}}$ which is related to the bulk infra red cut-off as $a={1}/{r_\infty}$.  

We may now employ our covariant holographic negativity construction given in Eq. (\ref {COHEN CONJ})
for the mixed state of adjacent intervals with lengths $l_1$ and $l_2$ in the holographic $\mathrm{\mathrm{\mathrm{CFT_{1+1}}}}$. This may be expressed as follows
\begin{equation}\label{nonextnegbulk}
\begin{aligned}
{\cal E}&=\frac{c}{8}\log\Bigg[\bigg(\frac{\beta_{+}}{\pi a}\bigg)\frac{\sinh{\big(\frac{\pi l_1}{\beta_{+}}\big)}\sinh{\big(\frac{\pi l_2}{\beta_{+}}\big)}}{\sinh\{{\frac{\pi}{\beta_{+}} (l_1+l_2)}\}}\Bigg]\\
&+\frac{c}{8}\log\Bigg[\bigg(\frac{\beta_{-}}{\pi a}\bigg)\frac{\sinh{\big(\frac{\pi l_1}{\beta_{-}}\big)}\sinh{\big(\frac{\pi l_2}{\beta_{-}}\big)}}{\sinh\{{\frac{\pi}{\beta_{-}} (l_1+l_2)}\}}\Bigg],\\
&=\frac{1}{2}({\cal E}_L+{\cal E}_R).
\end{aligned}
\end{equation}
Note that the Brown-Henneaux formula $c=\frac{3 R}{2 G_N^3}$ \cite{brown1986} has been used in the above expression. Interestingly we observe that the covariant holographic negativity
for the mixed state in question decouples into ${\cal E}_L$ and ${\cal E}_R$  with left and right moving inverse temperatures $\beta_{+}$ and $\beta_{-}$ respectively \cite{Zoltan}. 
We will have more to say about this issue in a later section.
%%%%%%%%%%%%%%%%%%%%%%%%%%%%%%%%%%%%%Extremal BTZ%%%%%%%%%%%%%%%%%%%%%%%%%%%%%%%%%%%%%%%%%%%%%%%%%%%%%%%%%%%%%%%%%%%%%%%%%%%%%%%%%%%%%%%%%%%%%%%%%%%%%%%%%%%%%%%%%%%%%%%%%%%%%%%%%%%%%%%%%%%%%%%%%%%%%%%%%%%
\subsection{Extremal rotating BTZ black hole}

We now turn our attention to the covariant holographic negativity for the mixed state of adjacent intervals in holographic $\mathrm{\mathrm{\mathrm{CFT_{1+1}}}}$s with a conserved charge, dual to bulk extremal rotating BTZ black holes. The extremal rotating BTZ metric is given as 
\begin{equation}\label{extbtz}
 ds^{2}=-\frac{(r^2-r_{0}^2)^2}{r^2}dt^2 + \frac{r^2 dr^2}{(r^2-r_{0}^2)^2}+ r^2(d\phi-\frac{r_{0}^2}{r^2}dt)^2.
\end{equation}
In the extremal limit $(r_+=r_-)$, the mass $M$ is equal to the angular momentum, $J$ $(M=J=2 r_0^2)$ and the corresponding Hawking temperature vanishes ($T_H=0$). As earlier, we first describe the computation for the length ${\cal L}_{\gamma}$ of the bulk geodesic anchored on a interval $\gamma$ of length $\Delta\phi$, in the rotating extremal BTZ geometry \cite{Hubeny:2007xt}.
To this end is required to perform the following coordinate transformations which map the extremal BTZ metric Eq. (\ref{extbtz}) to the $\mathrm{AdS_3}$ metric in the Poincar\'{e} coordinates (\ref{poincare}) 
\begin{equation}\label{extbtzmaps}
\begin{aligned}
&w_{+}=\phi+t-\frac{r_0}{r^2-r_0^2}, ~~~w_{-}=\frac{1}{2r_0}e^{2r_0(\phi-t)},\\
& ~~~~~~~~~~~~~~Z=\frac{1}{\sqrt{r^2-r_0^2}}e^{r_0(\phi-t)}.
\end{aligned}
\end{equation}
The geodesic length ${\cal L}_{\gamma}$ anchored on the interval $\gamma$ of length $\Delta\phi$ may now be expressed as follows
\begin{equation}\label{extlenght}
\begin{aligned}
 {\cal L}_{\gamma}&=\log\bigg[\frac{(\Delta \phi)^2}{ \varepsilon_{i}\varepsilon_{j}}\bigg],\\
\varepsilon_{i}&=\frac{1}{r_{\infty}}e^{r_0(\phi_{i}-t_0)}.
\end{aligned}
\end{equation}
In the original BTZ coordinates the geodesics length in Eq. (\ref {extlenght}) reduces to the following
\begin{equation}\label{s45}
{\cal L}_{\gamma}=\log\bigg[\frac{\Delta\phi}{a}\bigg]+\log\bigg[\frac{1}{r_0a}\sinh\big(r_0\Delta\phi\big)\bigg].
\end{equation}

It is now possible to employ our covariant holographic negativity construction for the mixed state of adjacent intervals with lengths $l_1$ and $l_2$ in the dual $\mathrm{\mathrm{\mathrm{CFT_{1+1}}}}$ as follows
\begin{equation}\label{extnegbulk}
\begin{aligned}
{\cal E}&=\frac{c}{8}\log\Bigg[\frac{l_1 l_2}{a(l_1+l_2)}\Bigg]\\
&+\frac{c}{8}\log\Bigg[\bigg(\frac{1}{r_0 a}\bigg)\frac{\sinh{\big(r_0l_1\big)}\sinh{\big(r_0l_2\big)} }{\sinh\{r_0(l_1+l_2)\}}\Bigg],
\end{aligned}
\end{equation}
where once again the Brown-Henneaux formula $c=\frac{3 R}{2 G_N^3}$ \cite{brown1986} has been used in the above expression and $a$ is the UV cut off in the $\mathrm{\mathrm{\mathrm{CFT_{1+1}}}}$.

Note that the first term in the previous expression resembles the ground state entanglement negativity of the mixed state in question whereas the second term is the entanglement negativity at an effective temperature given by the well known Frolov-Throne temperature $T_{\mathrm{FT}}=\frac{r_0}{\pi}$ \cite{PhysRevD.39.2125,Caputa:2013lfa}.
%%%%%%%%%%%%%%%%%%%%%%%%%%%%%%%%%%%%%Covariant Entanglement Negativity in a $\mathrm{\mathrm{\mathrm{CFT_{1+1}}}}$ dual to a Rotating BTZ Black Holes%%%%%%%%%%%%%%%%%%%%%%%%%%%%%%%%%%%%%%%%%%%%%%%%%%%%%%%%%%%
%%%%%%%%%%%%%%%%%%%%%%%%%%%%%%%%%%%%%%%%%%%%%%%%%%%%%%%%%%%%%%%%%%%%%%%%%%%%%%%%%%%%%%%%%%%%%%%%
%%%%%%%%%%%%%%%%%%%%%%%%%%%%%%%%%%%%%%%%%%%%%%%%%%%%%%%%%%%%%%%%%%%%%%%%%%%%%%%%%%%%%%%%%%%%%%%%
\section{Entanglement negativity for a $\mathrm{\mathrm{\mathrm{CFT_{1+1}}}}$ dual to a rotating BTZ black hole}\label{sec_CFT_rot_btz}
In this section we proceed to compute the entanglement negativity, through the replica technique, for the holographic $\mathrm{\mathrm{\mathrm{CFT_{1+1}}}}$ dual to non extremal and extremal rotating BTZ black holes to establish the veracity of our construction. To this end we first describe the computation of the entanglement negativity for such mixed states in a generic $\mathrm{\mathrm{\mathrm{CFT_{1+1}}}}$ given in \cite{Calabrese:2012ew,Calabrese:2012nk}. The entanglement negativity for the mixed state of adjacent intervals with lengths $l_1$ and $l_2$ obtained through the replica technique is related to a three-point function of the twist fields as follows\footnote{Note that in \cite{Calabrese:2014yza}, the authors study the entanglement negativity of a single interval with rest of the system  in a finite temperature $\mathrm{\mathrm{\mathrm{CFT_{1+1}}}}$ which requires a four point correlation of twist fields on an infinite cylinder. This is because of the complicated partial transpose over an infinite part of an infinite system at finite temperature. However this does not apply to the computation of the negativity for the mixed state of finite adjacent intervals which is still described by the three point function as given by eq.(\ref{negativity}) (See section {\bf 3} of \cite{Calabrese:2014yza}).}
 \begin{equation}\label{negativity}
{\cal E}=\lim_{n_e \to 1 }\ln \Big [\langle\mathcal{T}_{n_e}(-l_1)\overline{\mathcal{T}}^2_{n_e}(0)\mathcal{T}_{n_e}(l_2)\rangle \Big],
\end{equation}
where $n_e$ stands for even parity of $n$ and the limit $n_e\to 1$ 
is to be understood as an analytic continuation of even sequences of $n$ to $n_e =1$ \footnote{ Although a general construction for this analytic continuation is elusive it has been demonstrated by Calabrese et. al. in \cite {Calabrese:2012ew,Calabrese:2012nk,Calabrese:2009ez,Calabrese:2010he} for specific simple conformal field theories.}.
The scaling dimensions for the twist fields in the above equation are given as follows \cite{Calabrese:2012ew,Calabrese:2012nk}
\begin{equation}\label{scaling dimension}
\begin{aligned}
\Delta_{\mathcal{T}_{n_e}} =  \frac{c}{12}\Big(n_e - \frac{1}{n_e}\Big),\\
\Delta_{\mathcal{\bar{T}}_{n_e}^2} = \frac{c}{6}\Big(\frac{n_e}{2}- \frac{2}{n_e}\Big).
\end{aligned}
\end{equation}

Note that the holographic $\mathrm{\mathrm{\mathrm{CFT_{1+1}}}}$s with a conserved charge, dual to both the non-extremal and the extremal rotating BTZ black holes live on a twisted infinite cylinder . These $\mathrm{\mathrm{\mathrm{CFT_{1+1}}}}$s are related to those defined on the complex plane through the appropriate conformal maps as described below. Hence the three point function in 
Eq.( \ref{negativity}) for the entanglement negativity of the mixed state in question has to be evaluated on the corresponding twisted infinite cylinder. The transformation property of the three point in Eq. (\ref{negativity}) under a conformal map ($z\to w$)
is given as follows
\begin{equation}\label{three point}
\begin{aligned}
 \langle\mathcal{T}_{n_e}(w_1)\overline{\mathcal{T}}^2_{n_e}(w_2)\mathcal{T}_{n_e}(w_3)\rangle_{cyl} =\prod_i &\Big(\frac{dw_i}{dz_i}\Big)^{-h}\Big(\frac{d\bar{w}_i}{d\bar{z}_i}\Big)^{-\bar{h}}\\
 \langle\mathcal{T}_{n_e}(z_1,\bar{z_1})\overline{\mathcal{T}}^2_{n_e}(z_2,\bar{z_2})\mathcal{T}_{n_e}(z_3,\bar{z_3})\rangle_{\mathbb{C}},
\end{aligned}
\end{equation}
where $(z,\bar{z})$ are the coordinates on the plane and $(w,\bar{w})$ are the coordinates on the twisted cylinder. The mixed state of the two adjacent intervals of lengths $l_1$
 and $l_2$ may then be obtained by identifying the coordinates ($w_i$) on the cylinder as follows 
 \begin{equation}\label{config}
 w_1=\bar{w}_1=-l_1,~~~w_2=\bar{w}_2=0,~~~w_3=\bar{w}_3=l_2.
 \end{equation}
Once the three-point function on the cylinder is obtained as described above, the corresponding entanglement negativity for the adjacent intervals case may then be computed by utilizing Eq. (\ref{negativity}).
%%%%%%%%%%%%%%%%%%%%%%%%%%%%%%%%%%%Non-Extremal BTZ%%%%%%%%%%%%%%%%%%%%%%%%%%%%%%%%%%%%%%%%%%%%%%%%%%%%%%%%%%%
%%%%%%%%%%%%%%%%%%%%%%%%%%%%%%%%%%%%%%%%%%%%%%%%%%%%%%%%%%%%%%%%%%%%%%%%%%%%%%%%%%%%%%%%%%%%%%
\subsection{$\mathrm{\mathrm{CFT_{1+1}}}$ dual to the non-extremal rotating BTZ black hole}
We are now in a position to compute the entanglement negativity for the mixed state of adjacent intervals with lengths $l_1$ and $l_2$ in the finite temperature 
$\mathrm{\mathrm{\mathrm{CFT_{1+1}}}}$ dual to a non-extremal rotating BTZ black hole utilizing the replica technique. The  partition function $Z(\beta)$ for this dual finite temperature $\mathrm{\mathrm{\mathrm{CFT_{1+1}}}}$ with the angular momentum $J$ as the conserved charge are given as follows \cite{Hubeny:2007xt,Caputa:2013lfa}
\begin{equation}\label{s21}
 Z(\beta)=\mathrm{Tr}\big(e^{-\beta (H-i\Omega_{E}J )}\big)=\mathrm{Tr}\big(e^{-\beta_{+}L_0-\beta_{-}\bar{L}_0}\big),
\end{equation}
 where $\rho=e^{-\beta (H-i\Omega_{E}J )}$ is the corresponding density matrix, $J$ is the angular momentum, $\Omega_{E}=-i \Omega$ is the Euclidean angular velocity and the Hamiltonian is $H=L_0+\bar{L}_0$ with $L_0$ and $\bar{L}_0$ are the standard Virasoro zero modes. As stated earlier the non-extremal BTZ metric Eq. (\ref{nonextbtz}) is mapped to the Poincar\'{e} patch of $\mathrm{AdS_3}$ by the coordinate transformations Eq. (\ref{nonextmap}). Taking $r\gg r_{\pm}$ limit, one may determine the conformal map from the $\mathrm{\mathrm{\mathrm{CFT_{1+1}}}}$ on the complex plane to that on the twisted infinite cylinder as follows
\begin{equation}\label{non-extmaps}
\begin{aligned}
w(z)\equiv&e^{\frac{2\pi}{\beta_+}(\phi+i\tau)}=e^{\frac{2\pi}{\beta_+}z},\\
\bar{w}(\bar{z})\equiv&e^{\frac{2\pi}{\beta_-}(\phi-i\tau)}=e^{\frac{2\pi}{\beta_-}\bar{z}}.
\end{aligned}
\end{equation}
The entanglement negativity of the mixed state in question may then be obtained through Eqs. (\ref{three point}), (\ref{non-extmaps}) and (\ref{negativity}) which leads to the following expression 
\begin{equation}
\begin{aligned}
{\cal E}=&\frac{c}{8}\log\Bigg[\bigg(\frac{\beta_{+}}{\pi a}\bigg)\frac{\sinh{\big(\frac{\pi l_1}{\beta_{+}}\big)}\sinh{\big(\frac{\pi l_2}{\beta_{+}}\big)}}{\sinh\{{\frac{\pi}{\beta_{+}} (l_1+l_2)}\}}\Bigg]\\
&+\frac{c}{8}\log\Bigg[\bigg(\frac{\beta_{-}}{\pi a}\bigg)\frac{\sinh{\big(\frac{\pi l_1}{\beta_{-}}\big)}\sinh{\big(\frac{\pi l_2}{\beta_{-}}\big)}}{\sinh\{{\frac{\pi}{\beta_{-}} (l_1+l_2)}\}}\Bigg]+ \mathrm{constant},
\end{aligned}
\end{equation}
where $``\mathrm{constant}"$ term is non universal and depends on the full operator content of the $\mathrm{\mathrm{\mathrm{CFT_{1+1}}}}$ \cite{Calabrese:2012ew,Calabrese:2012nk}. It was described in \cite{Kulaxizi:2014nma,Jain:2017aqk} that the non universal constant may be neglected in the large central charge limit. As earlier the entanglement negativity may be expressed as
\begin{equation}
{\cal E}=\frac{1}{2}({\cal E}_L+{\cal E}_R).
\end{equation}
It is observed from the above expression that the universal part of the entanglement negativity for the $\mathrm{\mathrm{\mathrm{CFT_{1+1}}}}$ decouples into ${\cal E}_L$ and 
${\cal E}_R$ components with left and right moving inverse temperatures $\beta_{+}$ and $\beta_{-}$ respectively \cite{Zoltan}. The separation of the entanglement negativity into these two parts correspond to the two mutually commuting chiral components of the relevant Hamiltonian 
$H=L_0+ \bar {L}_0$ as described in \cite{Zoltan}. Remarkably, the universal part of the entanglement negativity for the mixed state of adjacent intervals computed through the replica technique exactly matches with the corresponding results Eq. (\ref {nonextnegbulk}) obtained from our covariant holographic negativity construction.

%%%%%%%%%%%%%%%%%%%%%%%%%%%%%%%%%%%%Extremal BTZ%%%%%%%%%%%%%%%%%%%%%%%%%%%%%%%%%%%%%%%%%%%%%%%%%%%%%%%%%%%%%%%%%%%%%%%%%%%%%%%%%%%%%%%%%%%%%%%%%%%%%%%%%%%%%%%%%%%%%%%%%%%%%%%%%%%%%%%%%%%%%%%%%%%%%%%%%%%%%%%%%%%%%%%%%%%%%%%%%%%%%%%%%%%%%%%%%%%%%%%%%%%%%%%%%%%%%%%%%%%%%%%%
\subsection{$\mathrm{\mathrm{CFT_{1+1}}}$ dual to the extremal rotating BTZ black hole}
As described earlier the extremal BTZ metric (\ref{extbtz}) is mapped to the Poincar\'{e} patch of $\mathrm{AdS_3}$ by the coordinate transformations  Eq. (\ref{extbtzmaps}). Note that as earlier the $\mathrm{\mathrm{\mathrm{CFT_{1+1}}}}$ dual to the extremal BTZ black hole is also defined on a twisted infinite cylinder. The corresponding conformal map to the complex plane may be determined for the asymptotic limit $(r\to \infty)$ of the coordinate transformations Eq. (\ref{extbtzmaps}) after a Wick rotation $t\to i\tau$ as follows
\begin{equation}\label{extmaps}
\begin{aligned}
w(z)\equiv&\phi+i\tau=z,\\
\bar{w}(\bar{z})\equiv&\frac{1}{2r_0}e^{2r_0(\phi-i\tau)}=\frac{1}{2 r_0}e^{2r_0\bar{z}}.
\end{aligned}
\end{equation}
 The entanglement negativity for the mixed state of adjacent intervals in the $\mathrm{CFT}_{1+1}$ dual to the extremal BTZ black hole may then be obtained from  Eq. (\ref{three point}), Eq. (\ref{extmaps}) and (\ref{negativity}) which now leads to the following expression
\begin{equation}
\begin{aligned}
{\cal E}&=\frac{c}{8}\log\Bigg[\frac{l_1 l_2}{a(l_1+l_2)}\Bigg]\\
&+\frac{c}{8}\log\Bigg[\bigg(\frac{1}{r_0 a}\bigg)\frac{\sinh{\big(r_0l_1\big)}\sinh{\big(r_0l_2\big)} }{\sinh\{r_0(l_1+l_2)\}}\Bigg] + \mathrm{constant}.
\end{aligned}
\end{equation}
It is important to note that the first term in the above expression is the entanglement negativity of the adjacent intervals for the $\mathrm{\mathrm{\mathrm{CFT_{1+1}}}}$ in its ground state whereas the second term describes the entanglement negativity at an effective Frolov-Thorne temperature $T_{FT}$ \cite{PhysRevD.39.2125,Caputa:2013lfa}. As explained earlier the $``\mathrm{constant}"$ term
is non universal and is sub leading in the large central charge limit 
\cite{Kulaxizi:2014nma,Jain:2017aqk}. Interestingly once again it is observed that the leading universal part of the entanglement negativity for the $\mathrm{\mathrm{\mathrm{CFT_{1+1}}}}$ obtained through the replica technique exactly matches in the large central charge limit with the holographic result in Eq. (\ref{extnegbulk}) .
Naturally this provides a strong consistency check for the universality of our construction in the $AdS_3/CFT_2$ scenario.

%%%%%%%%%%%%%%%%%%%%%%%%%%%%%%%%%%%%%%%%%%%%%%%%%%%%%%%%%%%%%%%%%%%%%%%%%%%%%%%%%%%%%%%%%%%%%%%%%%%%%%%%%%%%%%%%%%%%%%%%%%%%%%%%%%%%%%%%%%%%%%%%%%%%%%%%%%%%%%%%%%%%%%%%%%%%%%%%%%%%%%%%%%%%%%%%%%%%%%%%%%%%%%%%%%%%%%%%%%%%%%%%%%%%%%%%%%%%%%%%%%%%%%%%%%%%%%%%%%%%%%%%%%%%%%%%%%%%%%%
\section{Time dependent holographic entanglement negativity}\label{sec_tdhen}

In the previous section we obtained the holographic negativity for the mixed state of adjacent intervals in $\mathrm{\mathrm{\mathrm{CFT_{1+1}}}}$s dual to bulk stationary $\mathrm{AdS_3}$ geometry described by the rotating BTZ black holes. Note however that the full import of the covariant holographic negativity construction requires its application to time dependent mixed states in $\mathrm{\mathrm{\mathrm{CFT_{1+1}}}}$. To this end in this section we utilize our construction to compute the  holographic entanglement negativity for time dependent mixed state configurations of adjacent intervals in a $\mathrm{\mathrm{\mathrm{CFT_{1+1}}}}$ dual to a bulk Vaidya-$\mathrm{AdS_3}$ space time. This bulk space time being one of the simplest examples of a time-dependent gravitational configuration in $\mathrm{AdS_3}$, describes the collapse process and subsequent black hole formation.

The Vaidya-$\mathrm{AdS_3}$ metric in the Poincar\'e coordinates is given as \cite{Hubeny:2007xt,AbajoArrastia:2010yt}
\begin{equation}
ds^2=-\left[r^2 -m(v)\right]~dv^2 +2~dr~dv +r^2 ~dx^2,
\end{equation}
where $m(v)$ is a function of the (light cone) time $v$. If the mass function $m(v)$ is a constant then this space time describes a non-rotating BTZ black hole.

The space like geodesic anchored on an interval of length $l$ $( x\in[-l/2,l/2] )$ in the $\mathrm{CFT}_{1+1}$ dual to the bulk Vaidya-$\mathrm{AdS_3}$ configuration is described by the functions $r(x)$ and $v(x)$ with the following boundary conditions
\begin{equation}
r(-l/2)=r(l/2)=r_{\infty},~~~~~v(-l/2)=v(l/2)=v.
\end{equation}
Here, the UV cut-off $r_{\infty}\rightarrow \infty$ which is inversely related to the lattice spacing as $r_{\infty}=1/\varepsilon$. The length $\mathcal{L}$ of such a space like geodesic  may be expressed as follows
\begin{equation}
\mathcal{L}=\int_{-l/2}^{l/2}\sqrt{r^2+2 ~\dot{r}~\dot{v}-f(r,v)~\dot{v}^2},
\end{equation}
where the $\dot {v}$ and $\dot {r}$ represent the derivative of the functions $v$ and $r$ w.r.t the coordinate $x$. In order to compute the explicit form of the geodesic length $\mathcal{L}$, we assume that the mass function $m(v)$ is a slowly varying function of $v$ and $\dot{m}(v)\ll 1$ in an adiabatic approximation as described in \cite{Hubeny:2007xt}. The length $\mathcal{L}$ of the space like geodesic described above may be written as \cite{Hubeny:2007xt}
\begin{equation}
\mathcal{L}=\mathcal{L}_{reg}+ 2\log(2 ~r_{\infty}),
\end{equation}
where the finite part of the length $\mathcal{L}_{reg}$ of the space like geodesic is given as 
\begin{equation}\label{lreg}
\mathcal{L}_{reg}=\log\frac{\sinh^2\left(\sqrt{m(v)~l/2}\right)}{m(v)}.
\end{equation}
The covariant holographic entanglement negativity for the mixed state of adjacent intervals of lengths $l_1$ and $l_2$ may then be computed by utilizing our holographic construction Eq. (\ref{COHEN CONJ}). The finite part of the holographic entanglement negativity $\mathcal{E}_{fin}$ may then be expressed as follows
\begin{equation}\label{Efin}\small
\begin{aligned}
\mathcal{E}_{fin}=\frac{3}{ 16 G_N^{(3)}}\log\bigg[\frac{\sinh^2\left(\sqrt{m(v)~l_1/2~}\right)~\sinh^2\left(\sqrt{m(v)~l_2/2~}\right)}{\sinh^2\left(\sqrt{m(v)~(l_1+l_2)/2~}\right)m(v)}\bigg].
\end{aligned}
\end{equation}
The finite part $\mathcal{E}_{fin}$ in the above expression is obtained through the cut-off regularization i.e by subtracting the divergent part given below, from the full expression for the holographic entanglement negativity 
\begin{equation}\label{Ediv}
 \mathcal{E}_{div}=\frac{3}{ 8 G_N^{(3)}} \log[2 ~r_{\infty}].
\end{equation}
The characteristic behavior of the finite holographic entanglement negativity $\mathcal{E}_{fin}$ as a function of $v$ is depicted in fig. (\ref{f1}) with the choice of the mass function as $m(v)=\tanh v$ as described in \cite{Hubeny:2007xt,AbajoArrastia:2010yt}. Interestingly we observe that the absolute value of $\mathcal{E}_{fin}$ decreases monotonically with $v$ and saturates to a fixed value for large $v$ describing the thermalization of the mixed state in question. Note that this corresponds to the black hole formation in the bulk $\mathrm{AdS_3}$ geometry.
\begin{figure}[ht!]
 \centering
\includegraphics[scale=.5]{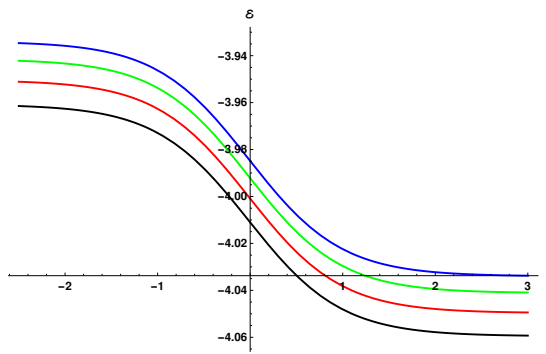}\label{f1}
 \caption  {$\mathcal{E}$ as a function of $v$ is plotted for $(l_1, l_2)=(.50, .60),~(.49,.61),~(.48,.62)$ and $(.47,.63)$, which are shown by blue, green, red and black curve respectively.}
 \label{f1}
\end{figure}

Observe that the above plot is in contrast to that described in \cite{Hubeny:2007xt} for the holographic entanglement entropy involving the bulk Vaidya-$AdS_3$ geometry. The authors there demonstrated that the holographic entanglement entropy increases monotonically and saturates to the value of the thermal entropy for large $v$. It is a general expectation from quantum information theory that an entanglement measure has to be non-negative and that it is a monotonically decreasing function of the temperature \cite {RevModPhys.75.715}. Hence, entanglement negativity being an upper bound on the distillable entanglement for mixed quantum states is expected to decrease monotonically during the process of thermalization.  On the other hand, the entanglement entropy receives contributions from both quantum and thermal correlations at finite temperatures and is dominated by the thermal correlations at high temperatures. As a result it increases monotonically as the mixed state thermalizes and saturates to the value of thermal entropy for large $v$.

Very interestingly, it has also been shown in the context of $(1+1)$ dimensional quantum field theories at zero temperature, that the renormalized entanglement entropy ($S_A^{ren}$) defined below, monotonically decreases along the renormalization group flow and is an example of an entropic $c$-function \cite{Casini:2004bw} 
\begin{equation}
 S_A^{ren}=l~\frac{\partial S_A(l)}{\partial l}.
\end{equation}
Here, $l$ represents the length of an interval $A$. Such a well defined renormalized entanglement entropy is  also expected to decrease monotonically under the renormalization group flow in higher dimensions \cite{Myers:2010tj,Liu:2012eea} (See \cite{Nishioka:2018khk} for a recent review).

Observe from fig.(\ref{f1}) that $\mathcal{E}_{fin}$ is negative. 
The finite part of the entanglement negativity having negative values  seems to be arising due to the cut-off regularization that we have utilized. It has been suggested in \cite{Banerjee:2015coc}  that a well defined renormalized entanglement negativity may correspond to a generalized $c$-function which is non-negative and monotonically decreases with temperature from its UV (Low temperature) value to IR (High temperature) value. Hence, it follows that such an expression for the renormalized entanglement negativity decreases monotonically during the process of thermalization.
Following \cite{Casini:2004bw,Banerjee:2015coc}, for the case of the adjacent intervals of equal length $l_1=l_2=l$, the renormalized entanglement negativity may be defined as
\begin{equation}
 {\cal E}_{ren}=l~\frac{\partial{\cal E}}{\partial l}
\end{equation}
\begin{figure}[ht!]
\centering
\includegraphics[scale=.5]{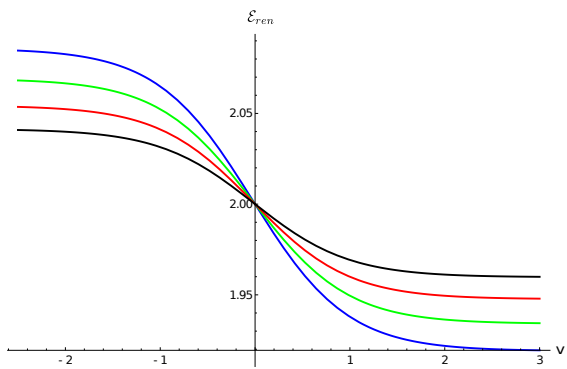}\label{f2}
 \caption{ $l~({\partial\mathcal{E}}/{\partial l})$ as a function of  $v$ is plotted for $l_1=l_2=l=.5,.45,.40$ and $.35$, which are shown by blue, green , red and black curve respectively.}
 \label{f2}
\end{figure}
Upon utilizing the length of the geodesics our  construction given in Eq. (\ref{COHEN CONJ}) we may obtain the renormalized holographic entanglement negativity as
\begin{equation}
\begin{aligned}
{\cal E}_{ren} =\frac{3 }{ 8 G_N^{(3)}} \left[l\sqrt{\tanh (v)} \coth \left(\frac{l}{2}  \sqrt{\tanh (v)}\right)\right]\\
-\frac{3 }{ 8 G_N^{(3)}} \left[l\sqrt{\tanh (v)} \coth \left(l \sqrt{\tanh (v)}\right)\right].
\end{aligned}
\end{equation}
The characteristic behavior of the renormalized holographic entanglement negativity ($\mathcal{E}_{ren}$) as a function of $v$ is depicted in fig. (\ref{f2}). Observe that as expected it is non-negative and a monotonically decreasing function of $v$. This  clearly indicates that the renormalized holographic entanglement negativity decreases monotonically and saturates to a fixed value as the mixed state in question thermalizes.

%%%%%%%%%%%%%%%%%%%%%%%%%%%%%%%%%%%Summary and Discussions%%%%%%%%%%%%%%%%%%%%%%%%%%%%%%%%%%%%%%%%%%%%%%%%%%%%%%%%%%%%%%%%%%%%%%%%%%%%%%%%%%%%%%%%%%%%%%Summary and Discussions%%%%%%%%%%%%%%%%%%%%%%%%%%%%%%%%%%%%%%%%%%%%%%%%%%%%%%%%%%%
\section{Summary and discussions}\label{sec_sumandcon}

To summarize we have advanced a covariant holographic entanglement negativity construction for mixed states of adjacent intervals in $(1+1)$-dimensional conformal field theories dual to non-static bulk 
$\mathrm{AdS_3}$ configurations. Our construction involves an algebraic sum of the lengths of bulk extremal curves anchored on the respective intervals. It is observed that the covariant holographic entanglement negativity reduces to the holographic mutual information between the corresponding intervals (in the large central charge limit of the $\mathrm{\mathrm{\mathrm{CFT_{1+1}}}}$). Interestingly similar result for this configuration has been obtained in \cite {Coser:2014gsa,Wen:2015qwa} in the context of global and local quenches in two dimensional conformal field theories.

To exemplify our construction and for a consistency check we have computed the entanglement negativity for the mixed state of adjacent intervals in $\mathrm{\mathrm{\mathrm{CFT_{1+1}}}}$s with a conserved charge, dual to the bulk stationary configurations described by non-extremal and extremal rotating BTZ black holes. For the finite temperature 
$\mathrm{\mathrm{\mathrm{CFT_{1+1}}}}$ dual to the bulk non-extremal rotating BTZ black hole we demonstrated that the covariant holographic entanglement negativity decouples into two components involving corresponding {\it left and right moving ``temperatures"}. For the extremal rotating BTZ case on the other hand, the covariant entanglement negativity reduces to a sum of the negativity for the ground state dual to a bulk AdS vacuum and that of an {\it effective ``thermal" state} characterized by a bulk {\it Frolov-Thorne temperature}.

To compare our holographic results we have computed the entanglement negativity through the replica technique for the dual $\mathrm{\mathrm{\mathrm{CFT_{1+1}}}}$s with a conserved charge.
Remarkably, in both cases we are able to demonstrate that the covariant holographic entanglement negativity exactly reproduces the universal part of the corresponding replica technique results in the large central charge limit of the $\mathrm{\mathrm{\mathrm{CFT_{1+1}}}}$. 

Furthermore to illustrate the time dependent aspect of our covariant holographic negativity construction we consider mixed states of adjacent intervals in a $\mathrm{\mathrm{\mathrm{CFT_{1+1}}}}$ dual to a bulk
Vaidya-$\mathrm{AdS_3}$ geometry. Very interestingly we are able to show from our construction that the renormalized holographic entanglement negativity monotonically decreases with time and saturates to a small fixed value indicating the thermalization of the mixed state in question. As is well known this corresponds to the black hole formation in the bulk $\mathrm{AdS_3}$ geometry and is completely consistent with quantum information theory expectations for this process\cite {RevModPhys.75.715}.

Naturally, these examples provide strong consistency checks for our proposed construction. The above results obtained utilizing our construction in the $\mathrm{AdS_3/CFT_2}$ scenario strongly motivate us to propose the generalization of our construction to higher dimensions in the spirit of \cite{Jain:2017xsu}.  
Following \cite {Hubeny:2007xt}, this leads us to generalize our covariant holographic entanglement negativity  construction for the mixed state of adjacent subsystems in $\mathrm{CFT_{d}}$s dual the non-static bulk $\mathrm{AdS_{d+1}}$ configurations in terms of the areas of the extremal surfaces instead of the static minimal surfaces. This is expressed as follows
\begin{equation}\label{COHEN CONJ hd} 
  {\cal E} = \frac{3}{16G_{N}^{(d+1)}} \Big[{\cal Y}_{A_1^t}^{ext}+{\cal Y}_{A_2^t}^{ext}-{\cal Y}^{ext}_{A_1^t\cup A_2^t}\Big],
\end{equation}
where ${\cal Y}_{\gamma}^{ext}$ denotes the bulk extremal surface anchored on the corresponding subsystem $\gamma$. However, it would require us to put our construction through explicit checks for specific higher dimensional examples and provide a plausible proof. We leave these significant open issues for future investigations.

Our covariant holographic entanglement negativity construction provides an ingenious prescription to determine the entanglement negativity of time dependent mixed states in generic holographic CFTs. Naturally, this will have significant implications for time dependent entanglement issues such as entanglement evolution in quantum quenches and  thermalization, quantum phase transitions and topological order in condensed matter physics. Our construction also promises significant insights into black hole formation and collapse scenarios, information loss paradox, unitarity restoration and the related firewall problem in the $\mathrm{AdS_3/CFT_2}$ context. We hope  to address these fascinating open issues in the near future.

%%%%%%%%%%%%%%%%%%%%%%%%%%%%%%%%%%%%%%%%%%%%%%%%%%%%%%%%%%%%%%%%%%%%%%%%%%%%%%%%%%%%%%%%%%%%%%
%%%%%%%%%%%%%%%%%%%%%%%%%%%%%%%%%%%%%%%%%%%%%%%%%%%%%%%%%%%%%%%%%%%%%%%%%%%%%%%%%%%%%%%%%%%%%%
%%%%%%%%%%%%%%%%%%%%%%%%%%%%%%%%%%%%%%%%%%%%%%%%%%%%%%%%%%%%%%%%%%%%%%%%%%%%%%%%%%%%%%%%%%%%%%
\section{Acknowledgment}
Parul Jain would like to thank Prof. Mariano Cadoni for his guidance and 
the Department of Physics, Indian Institute of Technology Kanpur, India for their warm hospitality. Parul Jain's work is financially supported by Universit\`a di Cagliari, Italy and INFN, Sezione di Cagliari, Italy.
%\newpage
%%%%%%%%%%%%%%%%%%%%%%%%%%%%%%%%%%%%%%%%%%%%%%%%%%%%%%%%%%%%%%%%%%%%%%%%%%%%%%
\bibliographystyle{utphys}
\bibliography{CoHEN}

\end{document}